\documentclass[twocolumn,floatfix,prl,aps,showpacs,superscriptaddress]{revtex4}
\usepackage{epsfig,graphicx,tabularx}
\usepackage{subfigure}
\usepackage{amsmath,amsfonts,amssymb}
\usepackage{color,braket}

\newcommand{\be}{\begin{equation}}
\newcommand{\ee}{\end{equation}}

\begin{document}

\title{Out-of-equilibrium states and quasi-many-body localization in
  polar lattice gases} 
\author{L. Barbiero} 

\affiliation{Dipartimento di Fisica e Astronomia "Galileo Galilei",
  Universit\`a di Padova, 35131 Padova, Italy}
\author{C. Menotti} \affiliation{INO-CNR BEC Center and
  Dipartimento di Fisica, Universit\`a di Trento, 38123 Povo, Italy}
\author{A. Recati} \affiliation{INO-CNR BEC Center and Dipartimento di
  Fisica, Universit\`a di Trento, 38123 Povo, Italy}
  \affiliation{Technische Universit\"at M\"unchen,  James-Franck-Strasse 1, 85748 Garching, Germany}
\author{L. Santos} \affiliation{Institut f\"ur Theoretische Physik,
  Leibniz Universit\"at Hannover, Appelstr. 2, DE-30167 Hannover,
  Germany}

\begin{abstract}
The absence of energy dissipation leads to an intriguing out-of-equilibrium dynamics for ultra-cold polar gases in optical lattices, characterized by 
the formation of dynamically-bound on-site and inter-site clusters of two or more particles, 
and by an effective blockade repulsion. These effects combined with the controlled preparation of initial states available in cold gases experiments 
can be employed to create interesting out-of-equilibrium states. These include
quasi-equilibrated effectively repulsive 1D gases for attractive
dipolar interactions and dynamically-bound crystals.  Furthermore,
non-equilibrium polar lattice gases can offer a promising scenario
for the study of quasi-many-body localization in the absence of quenched
disorder.  This fascinating out-of-equilibrium dynamics for
ultra-cold polar gases in optical lattices may be accessible in
on-going experiments.
 \end{abstract}


\maketitle




Out-of-equilibrium dynamics of isolated quantum systems has recently
attracted a major interest~\cite{Polkovnikov2011,Cazalilla2011}, in
particular in the context of ultra-cold gases, where dissipation is
basically absent~\cite{Bloch2008}.
Non-equilibrium quantum dynamics  constitutes an
exciting new field, notably in what concerns many-body
localization~(MBL), i.e. localization in excited states of
interacting many-body systems~\cite{Nandkishore2014}. Recent
cold-gases experiments are starting to unveil the non-trivial physics
of MBL~\cite{Schreiber2015}.

Although MBL is typically discussed in the presence of disorder,
localization may occur in absence of it, as first
discussed for $^3$He diffusion in $^4$He
crystals~\cite{Mikheev1983,Kagan1984}. Beyond a critical
concentration, immobile $^3$He clusters could lead to percolation for
the remaining $^3$He atoms.  Quasi-MBL and glassy dynamics without disorder are attracting a growing attention, and various
mechanisms for localization and eventual delocalization have been
discussed~\cite{Carleo2012,Schiulaz2014,Schiulaz2014b,Grover2014,Hickey2014,DeRoeck2014,
  Yao2014,Papic2015}.

Meanwhile, experiments on magnetic
atoms~\cite{Griesmaier2005,Lu2011,Aikawa2012} and polar
molecules~\cite{Ni2008,Wu2012, Takekoshi2014} are starting to reveal
the fascinating physics of dipolar gases. These gases are markedly
different from their non-dipolar counterparts due to the long-range
anisotropic character of the dipole-dipole
interaction~(DDI)~\cite{Lahaye2009,Baranov2012}.  Polar gases 
in optical lattices~(OLs) offer exciting possibilities for the study
of lattice models~\cite{Baranov2012} and quantum magnetism~\cite{Yan2013,DePaz2013}.

\begin{figure}[t]
\begin{center}
\includegraphics[width=0.8\columnwidth]{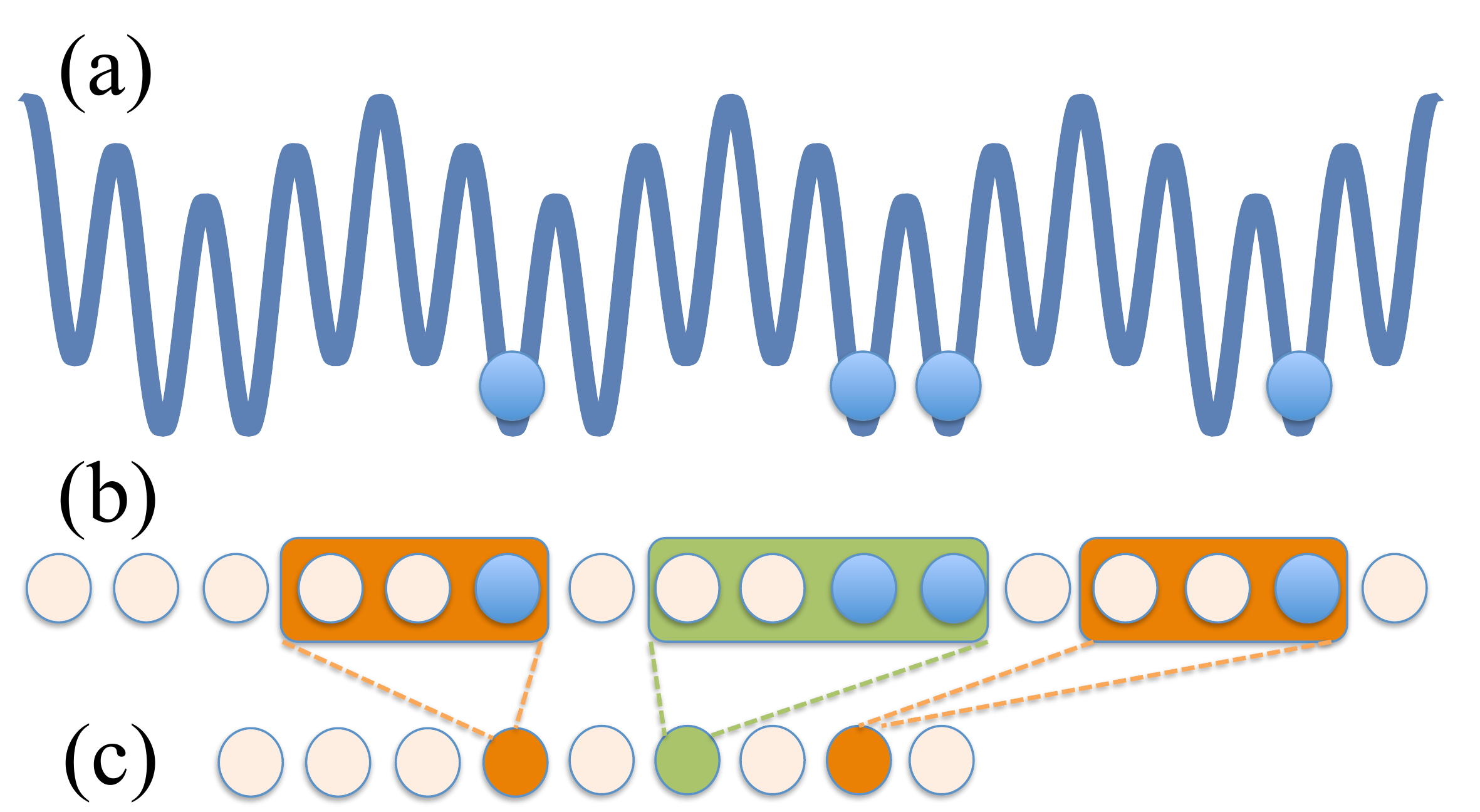}
\end{center}
\vspace*{-0.3cm}
\caption{(Color online) (a) A dimerized lattice can be used to create
  a gas of singlons and dynamically-bound NN dimers; (b) due to the BR
  a NN dimer~(singlon) forms a block $D\equiv 0011$~($S\equiv 001$);
  (c) effective lattice formed by the blocks D, S, and additional
  empty sites. This effective lattice is employed in
  model~\eqref{eq:Heff} to show the realization of quasi-MBL.}
\label{fig:1}
\end{figure}

In this Letter, we study non-equilibrium dynamics of 1D
polar lattice gases.  This dynamics is characterized by dynamically-bound
on-site and inter-site clusters (BCs) generalizing on-site repulsively-bound pairs in non-polar gases~\cite{hubbard,Winkler2006,Strohmaier2010}, and by blockade repulsion
(BR). We show how
these effects result in interesting
out-of-equilibrium states, including repulsive gases
with attractive DDI and dynamically-bound crystals.  Moreover, 
polar lattice gases allow for 
quasi-MBL without disorder, as we illustrate for
the set-up of Fig.~\ref{fig:1}.  These
scenarios can be realized in current experiments on polar
molecules in OLs.

\paragraph{Model.--}  We consider polar bosons in a 1D OL. 
For a deep lattice the system is described by the extended
Bose-Hubbard model~(EBHM)~\cite{Lahaye2009,Baranov2012}
\begin{equation}
H=-J\sum_{\langle ij \rangle} \hat b^\dagger_i \hat b_j+\frac{U}{2}\sum_i \hat n_i(\hat n_i-1)+V\sum_{i,r>0}\frac{\hat n_i \hat n_{i+r}}{r^3}, 
\label{EBH}
\end{equation}
where $\langle \cdots \rangle$ denotes nearest neighbor~(NN),
$b_i$~($b_i^{\dagger}$) destroys~(creates) bosons at the $i$-th
 site, $n_i=b_i^\dag b_i$, $J$ is the hopping rate, $U$
characterizes the combined on-site short-range interactions and DDI, and $V/r^3$ is the
strength of the DDI between sites placed $r\ge 1$ sites apart.  
$J$, $U$ and $V$ can be tuned independently 
by changing the lattice depth, the transverse confinement~\cite{Goral2003}, the orientation and strength of the polarizing field~($V<0$ for polarization along the lattice axis), 
and by Feshbach resonances. 

\begin{figure}[t]
\includegraphics[height=6cm,width=\columnwidth]{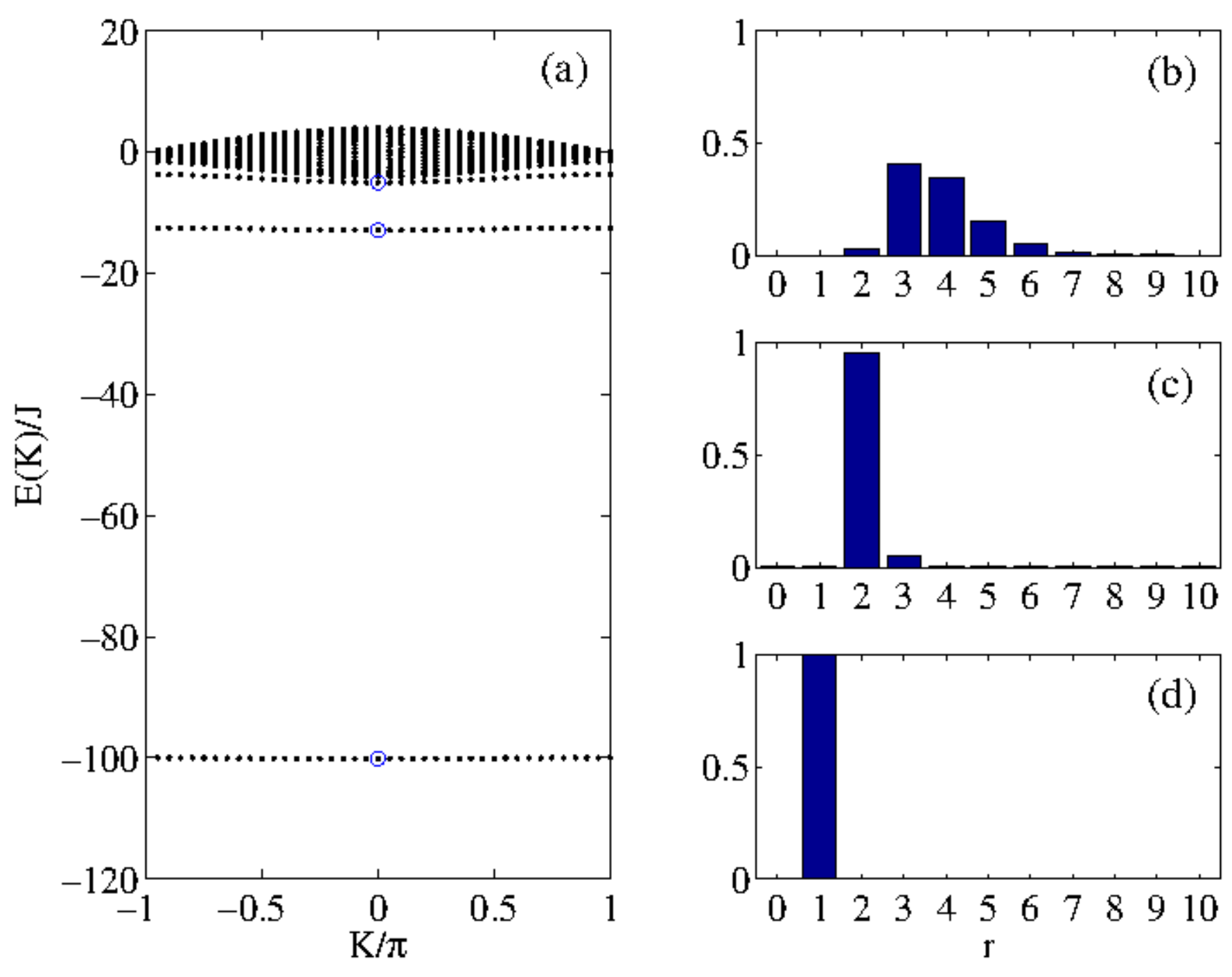}
\vspace*{-0.5cm}
\caption{(Color online) (a) Energy spectrum as a function of the
  center-of-mass quasi-momentum $K$ for two particles in $40$ sites
  for $U=0$ and $V=-100J$; density of the
  bound-states as a function of the relative coordinate $r$ for the
  three bound states at $K=0$, as marked by the blue circles in
  Fig.~(a), at energies $E/J \approx -5$ (b), $-13$ (c) and $ -100$
  (d). }
\label{fig:2}
\end{figure}



\paragraph{Bound pairs.--} 
We revisit first the concept of bound pairs in non-polar gases~($V=0$).  Doubly-occupied sites are
characterized by an interaction energy $U$.  If $|U|\gg J$, 
energy conservation maintains on-site pairs irrespective of the sign of
$U$~\cite{footnote-RBP}.  For $U>0$ those pairs, also called repulsively-bound pairs~\cite{Winkler2006}, 
are hence dynamically bound. Conversely, two separated
particles cannot be brought to the same site, i.e. singlons experience
hard-core repulsion~\cite{Mark-Haller2012}.  However,
singlons may resonantly move through on-site pairs
since a single-particle hopping swaps doublon and singlon positions~($21\to
12$)~\cite{Heidrich-Meisner2009,Ronzheimer2013}.
 
The long-range DDI allows for dynamically-bound inter-site 
pairs~\cite{Nguenang2009}.  Figure~\ref{fig:2}(a) depicts a typical
two-particle spectrum, for $U=0$ and $V=-100J$.  For each
center-of-mass quasi-momentum $K \in [-\pi,\pi]$, the spectrum
presents a continuum of scattering states and a discrete set of
isolated inter-site bound states~(BSs)~\cite{Valiente2010,footnote-RNN}, which as
for on-site bound pairs in non-polar gases, are maintained by energy
conservation, irrespective of the sign of $V$.
Figures~\ref{fig:2}(b-d) show the probability of finding two particles
$r$ sites apart for the BSs at $K=0$.  For binding energies close to
the continuum, the relative position of the pair delocalizes over many
sites~(Fig.~\ref{fig:2}(b)).  These delocalized BSs are for any
practical purposes indistinguishable from the scattering states.
Instead, as shown in Fig.~\ref{fig:2}(c-d), deeper BSs present a well defined relative distance $r$. 
Below we restrict the term bound pair~(BP) to deep BSs
at fixed $r\leq r_c$, where the critical $r_c$ is
defined as the largest $r$ satisfying the condition $f(r)=2
(J/V)^2/(r^{-3}-(r+1)^{-3})^2 \ll 1$~\cite{footnote-rc}.  Note 
that even if $U=0$, the inter-site DDI stabilizes an on-site BP that
is buried within the scattering states in Fig.~\ref{fig:2}(a), close to energies $E=U=0$. 
On-site and NN BPs demand $|V-U|\gg J$ to avoid resonances
between on-site and inter-site interactions~\cite{footnote-UV}.

\begin{figure*}[t]
\begin{center}
\includegraphics[width=2.0\columnwidth]{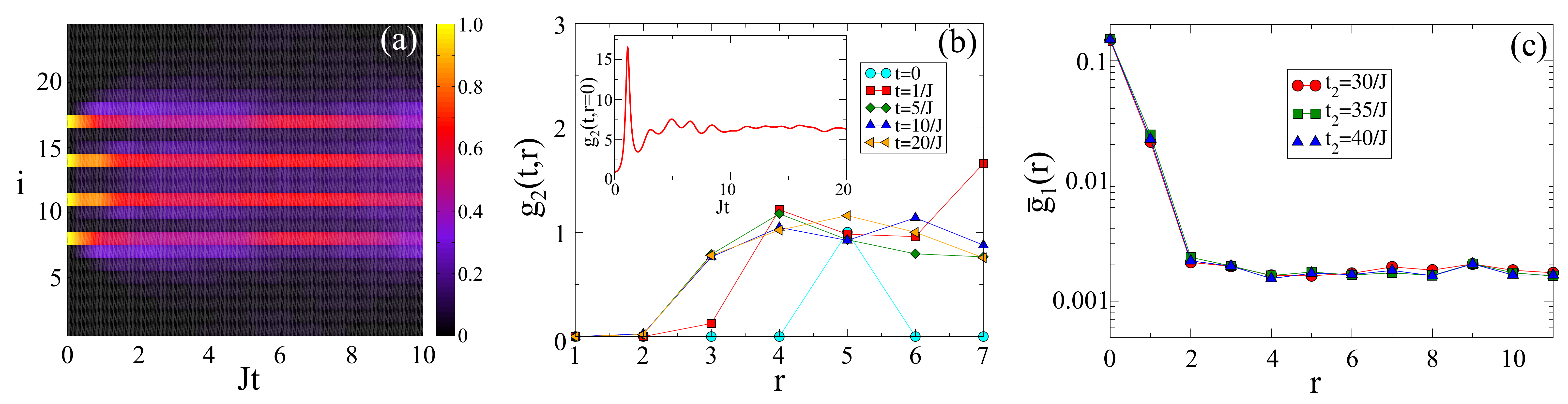}
\end{center}
\vspace*{-0.5cm}
\caption{(Color online) (a) \emph{Dynamically-bound crystal}: density $\langle n_i\rangle$~(obtained using t-DMRG)
for $U=0$ and $V=-100J$~($r_c=2$), for $4$ particles in $24$ sites initially placed $r_{in}=3$ sites apart. 
Note that since $r_{in}=r_c+1$, there is still some residual dynamics. For smaller $r_{in}\leq r_c$ the crystal is perfectly preserved in this time scale. 
(b)--(c) \emph{Effective repulsive 1D gas}: same parameters as in Fig.~(a)  but $r_{in}=5$:
(b)  $g_2(t,r>0)$ at different times (in the inset we show $g_2(t,r=0)$~\cite{footnote-homogeneous}) 
and (c) $\bar {g}_1(r)$, for $J (t_2-t_1)=5$ and different $t_2$ values. 
For both $\bar{g}_1$ and $g_2$, $r$ denotes the distance from an initially occupied site.
Note that not only the on-site density and density fluctuations 
converge to quasi-equilibrium, but also $\bar{g}_1$ and $g_2$ at longer distances.
Note as well 
that $g_2(t,0<r\leq r_c)=0$ due to BR.
 }
\label{fig:3}
\end{figure*}

\paragraph{Bound clusters.--} 
On-site interactions may bind more than two particles in on-site
BCs. However, on-site BCs are unstable against three-body losses and
play a relevant role only at relatively large lattice
fillings~\cite{footnote-DoublonClusters}.  Polar lattice gases allow
for inter-site BCs of more than two particles, each particle being 
within a distance $r\leq r_c$ of at least another
particle of the cluster.  Several important points must be
noted. First, although sites with more than one particle may be
involved, inter-site BCs are typically formed by singly-occupied
sites, and hence these clusters are in general stable against
three-body losses. Second, whereas on-site BPs are obviously precluded
for polarized Fermi gases, inter-site BPs and BCs are possible even in
that case. Third, in contrast to on-site BCs, inter-site BCs may
present internal resonances, e.g. the cluster $1101$ may remain
bound, but resonates with $1011$.  BCs are a general feature of
non-equilibrium polar lattice gases in any dimension even at low
fillings, as long as the DDI is large enough. In particular, massive
BCs of a size comparable to the whole system may be formed if the
mean-interparticle distance $R< r_c$.  An example of massive
BC is provided by particles initially placed at
regular distances $r_{in}\leq r_c$.  The absence of dissipation
maintains this \emph{dynamically-bound
crystal}~(Fig.\ref{fig:3}(a)).



\paragraph{Blockade repulsion.--} 
The formation of inter-site BPs has as a counterpart a vanishing
probability of finding the particles at a distance $r\leq r_c$ in
loose inter-site BSs and scattering states~($r_c=2$ for
Fig.~\ref{fig:2}(b)~\cite{footnote-rc}).  This
exclusion region leads to an effective BR between particles
initially at a distance $r_{in}>r_c$. This BR becomes evident
in the density-density correlation
$g_2(t,r)=\langle n_i(t)n_{i+r}(t)\rangle / ( \langle n_i(t)\rangle \langle n_{i+r}(t)\rangle)$. 
If $r_{in}>r_c$ at $t=0$, the subsequent dynamics shows BR,
i.e. $g_2(t>0,0<r\leq r_c)=0$, as discussed below.



\paragraph{Repulsive gas for attractive DDI.--} Combining BR with a proper initial-state 
preparation allows for the creation of a repulsive gas for attractive DDI.  Such a gas may be realized
by placing particles at the minima of a
superlattice with period $r_{in}>r_c$ and subsequently removing the
superlattice; atoms in sites with more than two particles may be eliminated by using resonant light~\cite{Sherson2010}.  Under these conditions no BP
or BC is present, and the system forms a singlon gas with
effective BR at radius $r_c$.  We have performed time-dependent density-matrix
renormalization group~(t-DMRG) simulations~\cite{Feiguin2005}  for $V=-100 J$ and $U=0$ ($r_c=2$) and $4$ particles
initially $r_{in}=5$ sites apart.  After a short time $\sim J^{-1}$,
the density $\langle n_j\rangle$ and $g_2(r=0)$ converge to the values expected for a homogeneous gas~(inset of Fig.~\ref{fig:3}(b))~\cite{footnote-homogeneous}. 
However, due to BR, $g_2(t,0<r\le r_c)=0$ for all $t$, whereas $g_2(t,r> r_c)$ has a
non-trivial dynamics reaching quasi-equilibrium~(Fig.~\ref{fig:3}(b)).
The time-averaged single-particle correlation $\bar
{g}_1(r)=\frac{1}{t_2-t_1}\int_{t_1}^{t_2} dt \langle b_i^\dag b_{i+r}
\rangle (t)$, shown for different times $t_{1,2}$ in
Fig.~\ref{fig:3}(c), also indicates quasi-equilibrium. The
equilibration of  $\bar{g}_1$ and $g_2$ relies on
the absence of BPs or BCs, contrasting with the quasi-MBL scenario 
below.

Although the effective repulsive 1D gas resembles a super-Tonks gas~\cite{Astrakharchik}, the physics behind is
very different.  In the super-Tonks case, an initially repulsive gas
is dynamically brought into an attractive regime. Even if in that
regime the two-body ground-state is a bound state, in absence of
dissipation the system remains in an excited state characterized by 
inter-particle repulsion.  In contrast, BR is crucially maintained by both the
absence of dissipation and by the lattice, which provides a finite band-
width and discrete particle motion.



\paragraph{Quasi-many-body localization.--} 
Polar lattice gases offer interesting possibilities for the study of
quasi-MBL without disorder.  BCs of $M$ particles move as a whole with 
hopping $J(J/V)^{M-1}$, and hence BCs with $M\gg 1$ are for any
practical purposes immobile~(although in-cluster
quasi-resonances may be still possible).  As for $^3$He~\cite{Mikheev1983,Kagan1984}, massive BCs
and BR may induce percolation for 
large-enough filling and $|V|/J$. Interestingly, as shown
below, 1D polar lattice gases may present quasi-MBL even for low fillings ($R\gg r_c$) and moderate DDI achievable in
experiments.

We illustrate the possibilities of 1D polar lattice gases for quasi-MBL within a simplified scenario. 
Resembling the recent experiment of Ref.~\cite{Schreiber2015} 
we consider a dimerized OL such that only the lower sites are populated~(Fig.~\ref{fig:1}(a)). After eliminating atoms in doubly-occupied sites~\cite{Sherson2010}, 
neighboring lower sites are hence either both occupied, or only
one of them, or none.  Then the superlattice is removed.
We consider $|V|/J$ such that $r_c=2$, and hence NN dimers form a BP, and a BR at $r_c=2$ is
established. As a result, blocks $0011$ and $001$ behave as well-defined particles that we call D and S,
respectively~(Fig.~\ref{fig:1}(b)). 
We neglect DDI for $r>r_c$ neighbors, since it is well within the bandwidth, and
obtain the effective model~\cite{footnote-SM}:
\begin{equation}
\hat H_{eff}=-\sum_{\langle ij \rangle} \left ( J \hat S_i^\dag \hat S_j +J_D \hat D_i^\dag \hat D_j+\Omega \hat D_{i}^\dag \hat S_j^\dag \hat S_{i}  \hat D_{j} \right ), 
\label{eq:Heff}
\end{equation}
where $i$, $j$ denote the sites of the effective lattice formed
by D's, S's, and empty sites of the original lattice belonging
neither to a D or an S~(Fig.~\ref{fig:1}(c)).  In $\hat H_{eff}$, $\hat
D_j$~($\hat S_j$) destroys a D~(S) at the effective site $j$.
Assuming for simplicity $U\gg V,J$, the D hopping rate is 
$J_D=\frac{8}{7}\frac{J^2}{V}$. The third term in $\hat H_{eff}$ is the swap
$DS\leftrightarrow SD$, occurring at rate
$\Omega=\frac{4}{3}\frac{J^2}{V}$. Note that a site
of the effective lattice is occupied by a D, an S, or empty. Due to
this hard-core constraint, in absence of swapping, D's and S's would
trivially localize each other, for any ratio $J_D/J$.

Swapping allows for the motion of D's and S's. However, the fact that $J_D,\Omega \ll J$ for $J/|V|\ll 1$
may result in localization following similar arguments as in Ref.~\cite{Schiulaz2014}. 
For $J/|V|\to 0$, the motion of S's is blocked by D's. 
For finite $J/|V|\ll 1$, the motion of D's changes the energy of the S gas~\cite{footnote-dE}. If this 
change, $\Delta E\gg J_D,\Omega$, the motion of D's is hindered. 
However, limited quasi-resonant D mobility, involving $\Delta E < J_D, \Omega$, remains possible, leading to partial D diffusion at times $\sim 1/\Omega$.

\begin{figure}[t]
\begin{center}
\includegraphics[width=0.9\columnwidth]{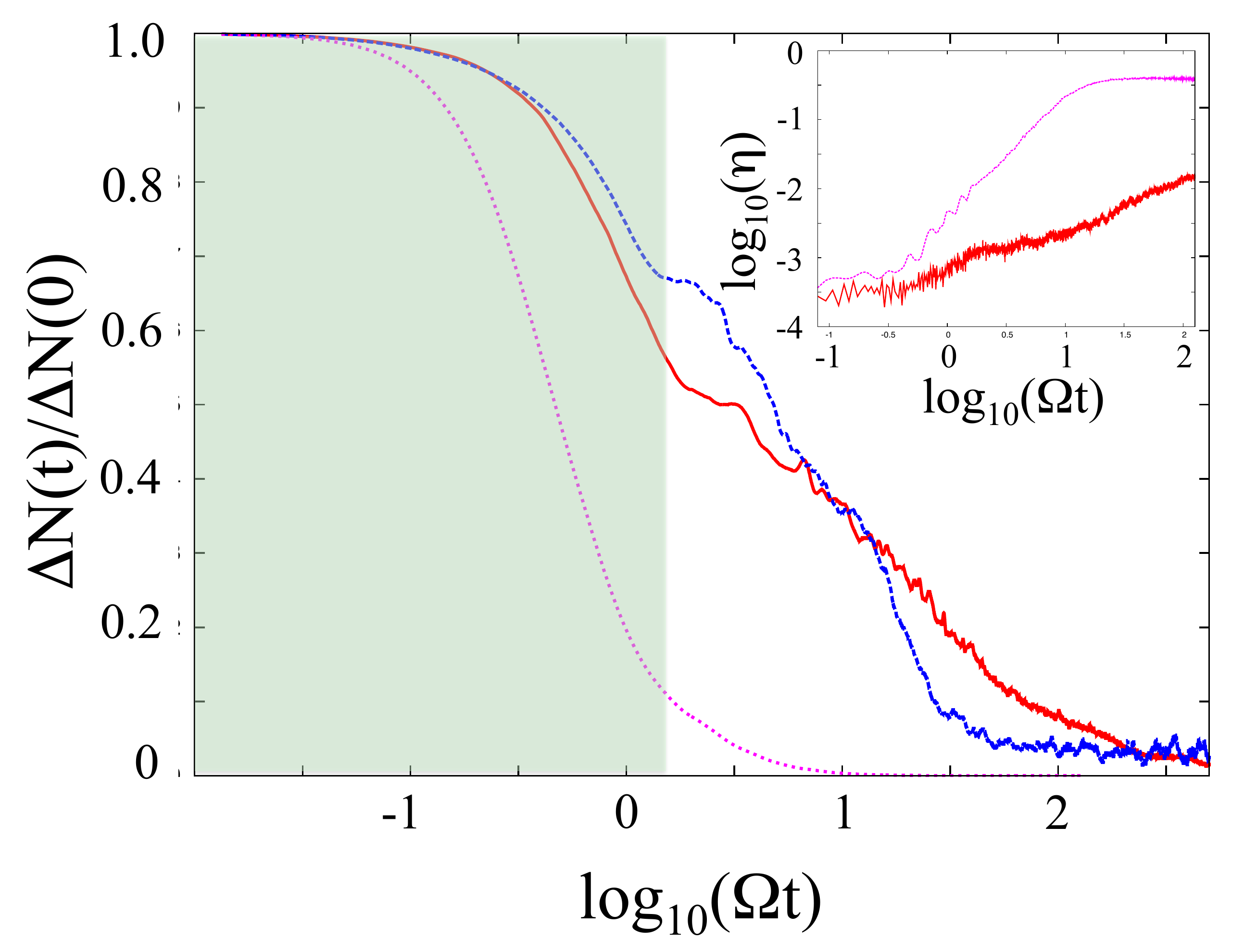}
\end{center}
\vspace*{-0.5cm}
\caption{(Color online) Inhomogeneity $\Delta N(t)$ as a function of
  $\Omega t$ for $\Omega/J=0.013$, for $(n_D,n_S,L)=(2,2,20)$~(dashed blue)
  and $(3,3,27)$~(solid red).  The results are obtained by exact
  diagonalization averaging over $50$ and $25$ random initial
  conditions, respectively. The shaded region indicates approximately
  the region of fast decay due to quasi-resonances. We depict for
  comparison the results for $(3,3,27)$ for $\Omega/J=0.13$~(dotted pink).  
  The inset shows the IPR for $(3,3,27)$ for 
  $\Omega/J=0.013$~(solid red) and $\Omega/J=0.13$~(dotted pink).}
\label{fig:4}
\end{figure}

We have performed exact diagonalization calculations with periodic
boundary conditions~(PBC) of the evolution of the many-body state $|\Psi(t)\rangle$ given by Model~\eqref{eq:Heff} for small
systems $(n_D,n_S,L)$ of $n_D$ D's, $n_s$ S's and $L$ lattice sites,
corresponding to $L_{eff}=L-3n_D-2n_S$ effective sites~\cite{footnote-SM}.
We average over various initial random distributions of D's and S's at
fixed positions in the effective lattice.  Figure~\ref{fig:4} shows for $\Omega/J=0.013$~($V=-100J$)
the dynamics of the inhomogeneity of D's, $\Delta N\equiv
\frac{1}{L_{eff}}\sum_{j=1}^{L_{eff}} |\langle \Psi(t) |\hat N_j -
\hat N_{j+1} |\Psi(t)\rangle |^2$ ~(with $\hat N_j\equiv \hat D_j^\dag
\hat D_j$)~\cite{Schiulaz2014}. Perfect homogeneity means $\Delta N=0$. 
In Fig.~\ref{fig:4} we depict for comparison the results for $\Omega/J=0.13$~\cite{footnote-V10}. 
Whereas for $\Omega/J=0.13$, D's diffuse
within a time scale $1/\Omega$, for $\Omega/J=0.013$ quasi-resonances allow
only a fraction of D's to delocalize in this time scale ~(shaded
region in Fig.~\ref{fig:4}) and a much slower dynamics follows. This
slow dynamics is characteristic of systems with PBC due the collective motion of all D's~\cite{Schiulaz2014b}.  Consistent with
this, the time scale of the slow D dynamics for the
$(n_D,n_S,L)=(3,3,27)$ case is approximately $10$ times longer than
that for $(2,2,20)$~\cite{footnote-slowdynamics}. We hence expect an exponentially diverging 
time scale for the slow-dynamics for growing number of dimers.

We may expand $|\Psi(t)\rangle=\sum_{\nu=1}^{n_{max}} \psi(\nu,t) | \nu \rangle$ over the $n_{max}=\binom{L_{eff}}{n_D+n_S}\binom{n_D+n_S}{n_S}$ 
many-body states $|\nu\rangle$ accounting for all possible distributions of S's and D's in the effective lattice.
MBL may be visualized as localization in this
many-body space.  The latter is best quantified by the inverse
participation ratio~(IPR), $\eta(t)\equiv\frac{1}{n_{max}}\left [
  \sum_\nu |\psi(\nu;t)|^4 \right ]^{-1}$; a fully
delocalized~(localized) state presents $\eta\sim
1$~($\sim1/n_{max}$). Whereas for $\Omega/J=0.13$, $\eta(t)\sim 1$  at $\Omega t < 10$, 
for $\Omega/J=0.013$, $\eta(t)$ remains very small even for $\Omega t \gg 1$~(inset of
Fig.~\ref{fig:4}), showing the appearance of quasi-localization in the
many-body space.



\paragraph{Experimental feasibility.--} 
The previous scenarios can be realized with polar molecules in OLs, as we illustrate for the case of NaK, which possesses an electric dipole of
$2.72$ Debye in its lowest ro-vibrational level~\cite{Park2015}. We consider the realistic case of partially polarized molecules with $1$ Debye. 
For a lattice spacing of $532$nm, $V/h\simeq 1$kHz. Assuming a
lattice depth of $18 E_{rec}$, with $E_{rec}/h\simeq 2.75$ kHz the
recoil energy, $J/h\simeq 10$Hz$=|V|/100$, and hence $1/\Omega\sim
1$s. As shown in Fig.~\ref{fig:4} the slow dimer dynamics can be much larger, 
stretching well beyond a minute, which is the typical maximal life time in experiments.
Localization can be explored either in expansion experiments, or by
site-resolved measurements.  Moreover, the formation of
a repulsive gas with attractive DDI can be readily monitored. 
Tightening an overall harmonic trap should result in the formation of an
incompressible crystalline core that can be revealed by measuring
the saturation of the mean radius of the sample and/or by site-resolved
measurements. Furthermore, BR hinders two or more molecules
to gather at the same site, preventing chemical recombination losses despite attractive DDI.



\paragraph{Summary.--} 
The absence of dissipation leads to rich out-of-equilibrium
dynamics in polar lattice gases characterized by the formation of
inter-site bound clusters and blockade repulsion even for attractive
DDI.  The combination of these effects with the control possibilities
of ultra-cold gases may allow the realization of effective
repulsive 1D gases with attractive DDI, the creation of
dynamically-bound crystals, and most interestingly, quasi-MBL in
absence of disorder. The latter opens interesting perspectives for observing a dynamical phase transition 
in polar lattice gases from a delocalized to a quasi-MBL regime as a function of the $V/J$ ratio.



{\it Acknowledgements.--} We thank X. Deng, P. Naldesi, H. Weimer, and A. Zenesini 
for discussions.  This work was supported by ERC~(QGBE grant), 
Provincia Autonoma di Trento, Cariparo Foundation~(Eccellenza grant 11/12), 
Alexander von Humboldt foundation, MIUR~(FIRB 2012, Grant No. RBFR12NLNA-002), cluster QUEST,
and DFG Research Training Group 1729. L. B. thanks the CNR-INO BEC Center in
Trento for CPU time.   L.S. thanks the BEC Center in Trento for its hospitality.



\newpage 

\section{Supplementary Material for "Out-of-equilibrium states and quasi-many-body localization in
  polar lattice gases"} 

We consider atoms occupying sparsely the lower sites of the 
lattice of Fig. 1(a) of the main text.  Neighboring lowest sites may
be both occupied, one occupied and one empty, or both empty. Atoms in
doubly-occupied sites or in sites with even higher occupation are
removed by resonant light.
As a result, when the additional superlattice is turned off, one
has either nearest-neighbor~(NN) dimers, or
singlons~(i.e. singly-occupied sites) that are at least three sites
apart from the nearest singlon or NN dimer. This provides our initial condition at $t=0$.

\begin{figure}[t]
\begin{center}
\includegraphics[width=\columnwidth]{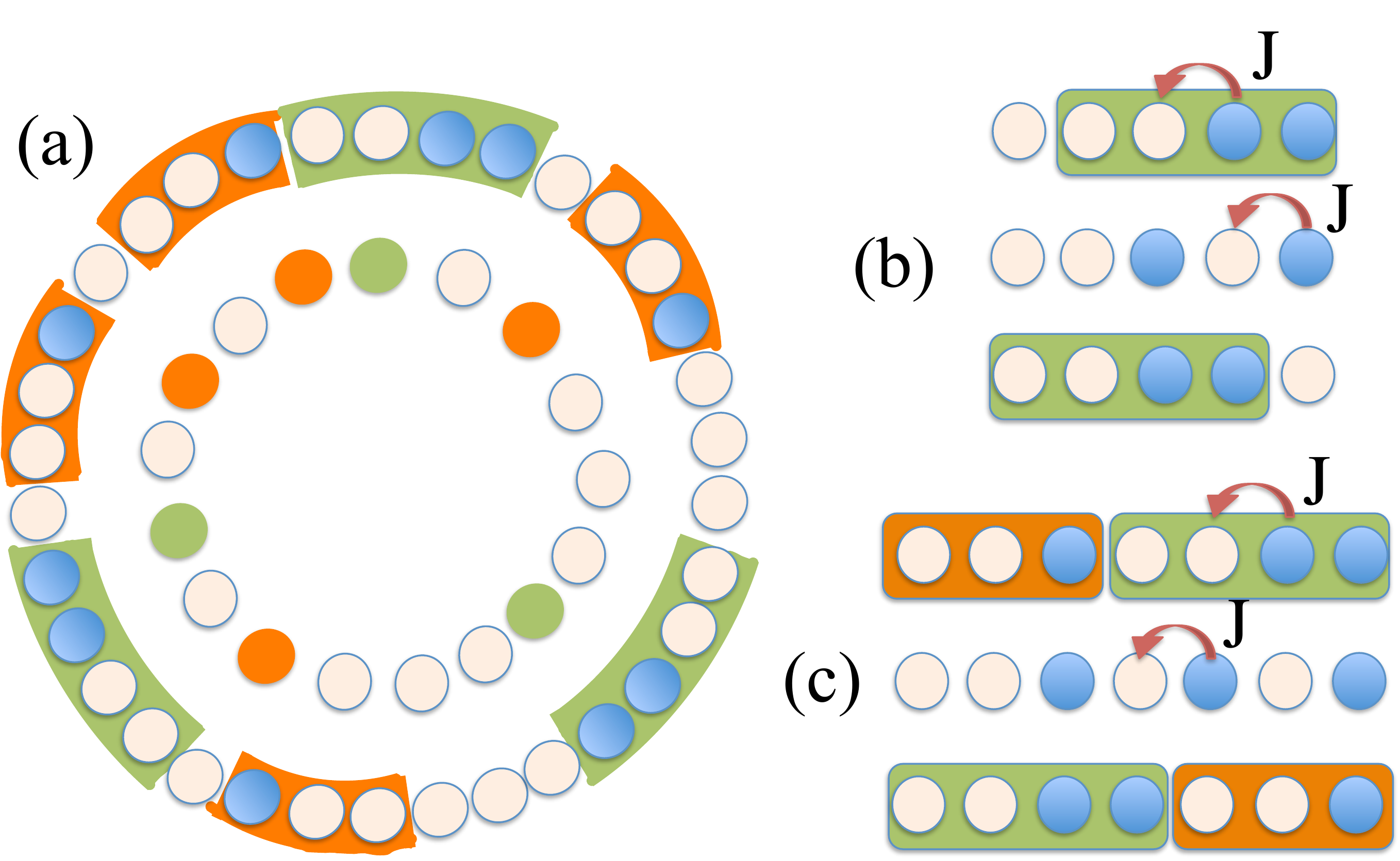}
\end{center}
\vspace*{-0.5cm}
\caption{(Color online) (a) The outer ring represents a 1D polar
  lattice gas with periodic boundary conditions, under the conditions
  discussed in the text. Singlons form effective clusters $S\equiv
  001$~(orange) and NN dimers effective clusters $D\equiv
  0011$~(green)~\cite{footnote-0}.  The inner ring represents the
  effective lattice corresponding to the real lattice in the outer
  ring . Each site of the effective lattice may be either
  empty~(white), occupied by an $S$~(orange), or by a
  $D$~(green). Figures (b) and (c) depict, respectively, the
  second-order processes responsible for the hopping of $D$'s into a
  neighboring empty effective site, and for $S$--$D$ swaps~(see
  text).}
\label{fig:S1}
\end{figure}

We assume a ratio $V/J$ such that $r_c=2$, ensuring that this minimal
distance persists during dynamics. Namely, due to blockade repulsion
no particle can ever be found at a distance equal or less than two
sites from a singlon or a NN dimer.  As a result each singlon and each
doblon have always two empty sites at their left and right. Hence we
can safely define the blocks $S\equiv 001$ for singlons and $D\equiv
0011$ for NN dimers~\cite{footnote-0}.  Due to periodic boundary
conditions, any possible state maps into a distribution of $S$'s and
$D$'s, and additional empty sites $0$, which do not belong to any $S$
or $D$. Therefore, we map the real system onto an effective lattice
with sites that are either empty, occupied by an $S$, or occupied by a
$D$~(see Fig.\ref{fig:S1}(a)).

The tunneling of $S$'s to empty effective sites takes place via
single-particle hopping $J$.  $D$'s can also move of one effective
site via second-order processes in two ways: (i) $011\to 020 \to 110$,
with an amplitude $J^2/(U-V)$; and (ii) $011\to 101 \to 110$~(see
Fig.~\ref{fig:S1}(b)) with an amplitude $8 J^2/7V$.  Hence the
doublon-hopping rate is $J_D=J^2/(U-V)+8 J^2/7V$. Assuming, for
simplicity, $U\gg V,J$, we approximate $J_D=8 J^2/7V$, as in the main
text.  Note however that this assumption is not strictly needed. It is
however necessary, as we mention in the main text, to fullfil the
condition $|U-V|\gg J$. Otherwise NN dimers become mobile at a hopping
rate similar to $J$ (exactly equal to $J$ in the case
$U=V$)~\cite{footnote-1}.
%
Finally, neighboring $D$'s and $S$'s may swap their position via
the second-order process sketched in Fig.~\ref{fig:S1}(c), with an amplitude
$\Omega=4J^2/3V$.


For $r_c=2$ and due to the $1/r^3$ decay of the dipole-dipole
potential, the interaction at distances larger than two sites is, by
definition of $r_c$, well within the band-width, and can hence be
neglected in a good approximation. As a result, the dynamics of the
polar lattice gas reduces to the hopping of $S$'s and $D$'s, and the
swap of $D$'s and $S$'s; this dynamics is described by model (2) of
the main text.

Note that this effective model resembles the one previously introduced
for binary mixtures of light and heavy
particles~\cite{Schiulaz2014,Schiulaz2014b,Grover2014} in the context
of MBL.  However in ~\cite{Schiulaz2014,Schiulaz2014b,Grover2014},
heavy particles block light particles, but light particles do not
block heavy ones. In model (2), D's block S's and vice versa, but S's
and D's can swap their positions. Note that, as mentioned in the main
text, these swaps are crucial, since without $S$-$D$ swaps (and as a
consequence of the hard-core constraint of the effective model), D's
and S's would block each other mutually, and the system would be
trivially localized. Indeed, it is the fact that the particles can in
principle extend over the whole lattice via $S$-$D$ swaps that makes
the problem non-trivial. As a consequence, the crucial parameter that
controls the quasi-many-body localization is $\Omega/J$, as shown in
Fig. 4 of the main text.

\bibliographystyle{prsty}

\begin{thebibliography}{99}

\bibitem{Polkovnikov2011} A. Polkovnikov, K. Sengupta, A. Silva, and
  M. Vengalattore, Rev. Mod. Phys. {\bf 83}, 863 (2011).

\bibitem{Cazalilla2011} See e.g. M. A. Cazalilla, R. Citro,
  T. Giamarchi, E. Orignac, and M. Rigol, Rev. Mod. Phys. {\bf 83},
  1405 (2011), and references therein.

\bibitem{Bloch2008} I. Bloch, J. Dalibard and W. Zwerger,
  Rev. Mod. Phys. {\bf 80}, 885 (2008).

\bibitem{Nandkishore2014} See e.g. R. Nankishore, and D. A. Huse,
  arXiv:1404.0686, and references therein.

\bibitem{Schreiber2015} M. Schreiber, S. S. Hodgman, P. Bordia, H. P. LŸ\"uschen, M. H. Fischer, R. Vosk, E. Altman, U. Schneider,
  and I. Bloch, arXiv:1501.05661.

\bibitem{Mikheev1983} V. A. Mikheev, V. A. Maidanov, and N. P. Mikhin,
  Solid State Commun. {\bf 48}, 361 (1983).

\bibitem{Kagan1984} Yu. Kagan and L. A. Maksimov,
  Zh. Eksp. Teor. Fiz. {\bf 87}, 348 (1984) [Sov. Phys. JETP {\bf 60},
    201 (1984)].

\bibitem{Carleo2012} G. Carleo, F. Becca, M. Schir\'o and M. Fabrizio,
  Scientific Reports {\bf 2}, 243 (2012).

\bibitem{Schiulaz2014} M. Schiulaz and M. M\"uller, AIP
  Conf. Proc. {\bf 1610}, 11 (2014).

\bibitem{Schiulaz2014b} M. Schiulaz, A. Silva, and M. M\"uller, Phys. Rev. B {\bf 91}, 184202 (2015).

\bibitem{Grover2014} T. Grover and M. P. A. Fisher,
  J. Stat. Mech. (2014) P10010.

\bibitem{Hickey2014} J. M. Hickey, S. Genway and J. P. Garrahan,
  arXiv:1405.5780 (2014).

\bibitem{DeRoeck2014} W. De Roeck and F. Huveneers, Phys. Rev. B {\bf
  90}, 165137 (2014).

\bibitem{Yao2014} N. Y. Yao, C. R. Laumann, J. I. Cirac, M. D. Lukin,
  and J. E. Moore, arXiv:1410.7407 (2014).

\bibitem{Papic2015} Z. Papic, E. M. Stoudenmire, Dmitry A. Abanin,
  arXiv: 1501.00477.

\bibitem{Griesmaier2005} A. Griesmaier, J. Werner, S. Hensler,
  J. Stuhler, and T. Pfau, Phys. Rev. Lett. {\bf 94}, 160401 (2005).

\bibitem{Lu2011} M. Lu, N. Q. Burdick, S. H. Youn, and B. L. Lev,
  Phys. Rev. Lett. {\bf 107}, 190401 (2011).

\bibitem{Aikawa2012} K. Aikawa, A. Frisch, M. Mark, S. Baier,
  A. Rietzler, R. Grimm, and F. Ferlaino, Phys. Rev. Lett. {\bf 108},
  210401 (2012).

\bibitem{Ni2008} K.-K. Ni et al, Science {\bf 322}, 231 (2008)

\bibitem{Wu2012} C.-H. Wu, J. W. Park, P. Ahmadi, S. Will, and
  M. W. Zwierlein, Phys. Rev. Lett. {\bf 109}, 085301 (2012).

\bibitem{Takekoshi2014} T. Takekoshi, L. Reichs\"ollner,
  A. Schindewolf, J. M. Hutson, C. R. Le Sueur, O. Dulieu,
  F. Ferlaino, R. Grimm, and H.-C. N\"agerl, Phys. Rev. Lett. {\bf
    113}, 205301 (2014).

\bibitem{Lahaye2009} See e.g. T. Lahaye, C. Menotti, L. Santos,
  M. Lewenstein, and T. Pfau, Rep. Prog. Phys. {\bf 72}, 126401
  (2009), and references therein.

\bibitem{Baranov2012} See e.g.  M. A. Baranov, M. Dalmonte,
  G. Pupillo, and P. Zoller, Chem. Rev. {\bf 112}, 5012 (2012), and
  references therein.

\bibitem{Yan2013} B. Yan, S. A. Moses, B. Gadway, J. P. Covey,
  K. R. A. Hazzard, A. M. Rey, D. S. Jin, and J. Ye, Nature {\bf 501},
  521 (2013).

\bibitem{DePaz2013} A. de Paz, A. Sharma, A. Chotia, E. Mar\'echal,
  J. H. Huckans, P. Pedri, L. Santos, O. Gorceix, L. Vernac, and
  B. Laburthe- Tolra, Phys. Rev. Lett. {\bf 111}, 185305 (2013).

\bibitem{hubbard} J. Hubbard, Proc. Roy. Soc. A {\bf 276}, 238 (1963).

\bibitem{Winkler2006} K. Winkler, G. Thalhammer, F. Lang, R. Grimm,
  J. Hecker-Denschlag, A. J. Daley, A. Kantian, H. P. B\"uchler, and
  P. Zoller, Nature {\bf 441}, 853 (2006).

\bibitem{Strohmaier2010} N. Strohmaier, D. Greif, R. J\"ordens,
  L. Tarruell, H. Moritz, T. Esslinger, R. Sensarma, D. Pekker,
  E. Altman, and E. Demler, Phys. Rev. Lett {\bf 104}, 080401 (2010).

\bibitem{Goral2003} K. G\'oral, L. Santos, and M. Lewenstein,
  Phys. Rev. Lett. {\bf 88}, 170406 (2002).

\bibitem{footnote-RBP} In 1D a bound pair appears for any finite $U$,
  but unless $|U|\gg J$ the pair remains very delocalized. Only for
  $|U|\gg J$ a deeply-bound on-site pair is formed.

\bibitem{Mark-Haller2012} M. J. Mark, E. Haller, K. Lauber,
  J. G. Danzl, A. Janisch, H. P. B\"uchler, A. J. Daley, and
  H.-C. N\"agerl, Phys. Rev. Lett. {\bf 108}, 215302 (2012).

\bibitem{Heidrich-Meisner2009} F. Heidrich-Meisner, S. R. Manmana,
  M. Rigol, A. Muramatsu, A. E. Feiguin, and E. Dagotto, Phys. Rev. A
  {\bf 80}, 041603 (R) (2009).

\bibitem{Ronzheimer2013} J. P. Ronzheimer, M. Schreiber, S. Braun,
  S. S. Hodgman, S. Langer, I. P. McCulloch, F. Heidrich-Meisner,
  I. Bloch, and U. Schneider, Phys. Rev. Lett. {\bf 110}, 205301
  (2013).


\bibitem{Nguenang2009} J.-P. Nguenang and S. Flach, Phys. Rev. A {\bf
  80}, 015601 (2009).

\bibitem{Valiente2010} M. Valiente, Phys. Rev. A {\bf 82}, 042102
  (2010).

\bibitem{footnote-RNN} We used
 a maximal range of the DDI of $R_{NN}=6$.
  Although this limits the maximal number of possible two-body bound
  states to $R_{NN}+1$, it does not
  affect the results under the conditions we analyze.

\bibitem{footnote-rc} Quantum deeply bound states differ from
  classical states of two particles at a fixed distance $r$ only by a
  perturbative correction. If $f(r)\ll 1$ the probability to find the particles at a distance different than $r$ is negligible. 
  In practice, we
  may introduce a small $\epsilon=0.01$ and define $r_c$ such
  that $f(r_c)\lesssim\epsilon$ but $f(r_c+1)\gg\epsilon$. For
  $V=-100J$~(Fig.~\ref{fig:2}) this criterion leads to $r_c=2$.


\bibitem{footnote-UV} Note that if $U=V$, a NN dimer resonates with an on-site pair. 
This allows the pair of particles to move resonantly along the lattice with the bare single-particle hopping.

\bibitem{footnote-DoublonClusters} This remains true for the formation
  of clusters of doubly-occupied sites, which may be maintained in short-range
  interacting gases due to second-order processes~\cite{Carleo2012}.

\bibitem{Sherson2010} J. F. Sherson, C. Weitenberg, M. Endres,
  M. Cheneau, I. Bloch, and S. Kuhr, Nature {\bf 467}, 68 (2010).
  
\bibitem{Feiguin2005} S. R. White and A. E. Feiguin, Phys. Rev. Lett. {\bf 93}, 076401 (2004); 
A. E. Feiguin and S. R. White, Phys. Rev. B {\bf 72}, 020404(R) (2005).

\bibitem{footnote-homogeneous} Since by construction the maximal occupation per site is one, $\langle n_i^2 \rangle=\langle n_i \rangle$, and 
$g_2(t,r=0)=\langle n_i \rangle^{-1}$. Hence, due to the homogenization of the density, $g_2(r=0)$ quickly converges to the inverse density. For the 
case of Fig.~\ref{fig:3}(b)~(4 particles in 24 sites), $g2(r=0)\to 6$. 

\bibitem{Astrakharchik} G. E. Astrakharchik, D. Blume, S. Giorgini,
  and B. E. Granger, Phys. Rev. Lett. {\bf 92}, 030402 (2004);
  M. D. Girardeau and G. E. Astrakharchik, Phys. Rev. Lett. {\bf 109},
  235305 (2012).

\bibitem{footnote-SM} See Supplemental Material.

\bibitem{footnote-dE}  For $J_D=\Omega=0$, immobile D's form effective box potentials, whose levels are occupied by the S's. For $J_D,\Omega>0$, the motion of D's 
changes the energies of the box levels and the distribution of S's among them.
As a result the energy of the system changes by an amount $\Delta E$.

\bibitem{footnote-V10} $\Omega/J = 0.13$ is considered just to illustrate better the
localization for $\Omega/J\ll 1$ in model~\eqref{eq:Heff}. Note that $\Omega/J =
0.13$ does not describe a polar gas at $V = -10J$, which has $r_c=1$,
and hence cannot be mapped to model~\eqref{eq:Heff}.

\bibitem{footnote-slowdynamics} Collective motion of two D's
  occurs in a time scale $\tau_2\sim \Delta E/\Omega^2$,
  where $\Delta E$ is the typical energy shift when displacing one
  D. In our simulations $\Omega\tau_2\simeq 10$. For the case of three
  D's the time scale for the collective motion is 
  $\tau_3 \sim \Delta E^2/\Omega^3$, and hence $\Omega\tau_3\simeq
  100$ in good agreement with our results.
  
\bibitem{Park2015} J. W. Park, S. A. Will, and M. W. Zwierlein, arXiv:1505.00473.

\end{thebibliography}

\begin{thebibliography}{}

\bibitem{footnote-0} Obviously we could have chosen $D\equiv 1100$ and
  $S\equiv 100$, and the mapping to the effective model would be
  identical.

\bibitem{footnote-1} Resonances between on-site interaction $U$ and NN
  interaction $V$ may lead to interesting physics, which lies beyond
  the scope of the present analysis.


\bibitem{Schiulaz2014} M. Schiulaz and M. M\"uller, AIP
  Conf. Proc. {\bf 1610}, 11 (2014).

\bibitem{Schiulaz2014b} M. Schiulaz, A. Silva, and M. M\"uller,
  Phys. Rev. B {\bf 91}, 184202 (2015).

\bibitem{Grover2014} T. Grover and M. P. A. Fisher,
  J. Stat. Mech. P10010 (2014).


\end{thebibliography}

\end{document}